\documentclass{llncs}
\usepackage{llncsdoc}

\usepackage{multirow}
\usepackage{courier}
\usepackage{amsmath}
\usepackage{amsfonts}
\usepackage{amssymb}
\usepackage{subfig}
\usepackage{enumitem}
\usepackage{graphicx}
\usepackage{multicol}
\usepackage[labelfont=bf]{caption}
\usepackage{subfig}
\usepackage{float}
\usepackage{color}
\usepackage{pifont}
\usepackage{algorithm}
\usepackage{algpseudocode}
\usepackage{subfig}
\usepackage{enumitem}
\usepackage{graphicx}
\usepackage{multicol}
\usepackage{amssymb}
\usepackage{hhline}
\usepackage[labelfont=bf]{caption}
\usepackage{subfig}
\usepackage{float}
\usepackage{color}
\usepackage{graphicx}
\usepackage{multirow}
\usepackage{tabularx}
\usepackage{array,graphicx}
\usepackage{booktabs}
\usepackage{pifont}
\usepackage[utf8]{inputenc}
\usepackage[english]{babel}
\usepackage{amsmath}

\usepackage{booktabs} 
\usepackage{amsmath}
\usepackage{subfig}
\usepackage{amssymb}
\usepackage{graphicx}
\usepackage{enumitem}
\usepackage{textcomp }

\makeatletter

\begin{document}
\title{Social Media and User Privacy}
\author{Ghazaleh Beigi}
\institute{\email{gbeigi@asu.edu} \\Arizona State University, Tempe, Arizona, USA}

\maketitle

\vspace{-10pt}
\begin{abstract}
	Online users generate tremendous amounts of data. To better serve users, it is required to share the user-related data among researchers, advertisers and application developers. Publishing such data would raise more concerns on user privacy. To encourage data sharing and mitigate user privacy concerns, a number of anonymization and de-anonymization algorithms have been developed to help protect privacy of users. This paper reviews my doctoral research on online users privacy specifically in social media. In particular, I propose a new adversarial attack specialized for social media data. I further provide a principled way to assess effectiveness of anonymizing different aspects of social media data. My work sheds light on new privacy risks in social media data due to innate heterogeneity of user-generated data.
\end{abstract}
\vspace{-25pt}
\section{Introduction}
Explosive growth of social media in the last decade has drastically changed the web and billions of people all around the globe can freely conduct numerous activities in a rich \textit{heterogeneous} environment. The resulted user-generated social media data consists of different aspects such as links, posts and profile information~\cite{halvarisnam2016,Beigi2018congruity}. This provides opportunities for researchers and business partners to study and understand individuals at unprecedented scales. However, publishing social media data risks exposing people's privacy as it is rich in content and relationship and contains individuals' sensitive information, resulting in privacy leakage~\cite{narayanan2009anonymizing}. For example, users' vacation plans, medical conditions and demographic information can be easily inferred from their posts.

Privacy issues of users mandate social media data publishers to protect users' privacy by anonymizing the data. One straightforward anonymization technique is to remove ``Personally Identifiable Information'' (a.k.a. PII) such as names, user ID, age and location information and keep the social graph structure as is. This solution has been shown to be far from sufficient to protects people's privacy~\cite{narayanan2009anonymizing}. Consequently, various protection techniques have been proposed for anonymizing each aspect of the heterogeneous social media data such as graph data structure~\cite{liu2008towards}, and users' textual information~\cite{TextAnonymization}. 

Existing anonymization techniques often make a specific assumption. In particular, these works assume that it's enough to anonymize each aspect of heterogeneous social media data independently. At the first glance, this assumption makes sense as anonymization takes time and effort. Moreover, users privacy is protected while the data utility is preserved at the highest possible level. For example, lets consider the simplest case study in which published data includes only two aspects such as (i) structural and (ii) textual (e.g., posts) information. We will then have options as shown in Table \ref{Tab:Hypothesis} to anonymize the data. To ensure anonymization efficiency, as each aspect can be of different data types, a common practice is to anonymize each aspect independently and thud case 4 is the backbone of the anonymization techniques for publishing data which is clearly the strongest protection of privacy. 

However, it is not clear whether latent relation between different aspects of the data could be captured and consequently sensitive information could be still leaked from the anonymized data considering each of these cases. In this research, we are interested to study these issues by answering the following questions:
\begin{itemize}[leftmargin=*]
	\item \textbf{(RQ1)}: Is the data private if just one of its two aspects is anonymized?
	\item \textbf{(RQ2)}: Is case 4, the strongest among four cases, sufficient for anonymizing social media data? 
\end{itemize} 
\begin{table}[t]\vspace{-0.20cm}
	\centering
	\small
	\caption{\textbf{Four different cases for social media data anonymization. Each check mark corresponds to the aspect of data being anonymized.}}
	\begin{tabular}{|l|l|l|l|l|}
		\hline
		& Case 1 & Case 2 & Case 3& Case 4\\ \hhline{=====}
		Structural Anonymization & \ding{55} & \ding{55} & \checkmark & \checkmark\\ 
		Textual Anonymization &  \ding{55} & \checkmark & \ding{55} & \checkmark\\ \hline
	\end{tabular}\label{Tab:Hypothesis}
	\vspace{-0.5cm}
\end{table}
We seek to answer these questions by taking an adversary approach to assay the privacy level of anonymized social media data. 
\vspace{-15pt}
\section{Proposed Adversarial Technique for Heterogeneous Social Media Data}
\vspace{-10pt}
Previous de-anonymization attacks require a list of target users as well as background knowledge for each targeted user. Target user is an individual with the known identity in social media and will be mapped to a user in the given anonymized data. This requires time and effort to find a proper set of target users and gather their knowledge which may not be realistic in practice. We propose a new generation of adversarial attacks specialized for social media data. This approach does not require background knowledge and publicly available information in social media are the adversaries' only source of information. The adversary can send queries to these APIs, anytime during the adversarial process. 

In this work we assume that data has two aspects, i.e., users' textual information $\mathcal{A}_1$ and graph structure $\mathcal{A}_2$ and the goal is to map user $\mathit{u}$ in anonymized dataset to a real profile in given social media $\mathcal{T}$. We posit that this attack is not limited to only two aspects or structural and textual information and could be generalized through abstraction to different social media data. The first step is to extract the most revealing information of $\mathit{u}$ according to her textual information. We use tf-idf scores to find the top-$k$ posts as the most revealing information. The second step includes selecting a set of candidate profiles for $u$ by searching for the extracted information from the previous step through $\mathcal{T}$'s search engine. Finally, the similarity between $u$ and her candidates are calculated using the combination of features of existing data components, structural and textual information. Degree distribution of $u$'s neighbors and textual vector of $u$'s posts are corresponding structural and textual feature, respectively. Features of the most similar users to $u$ (e.g., neighbors) are also incorporated as well to reduce the anonymization effect while capturing the hidden relation between different aspects of the data. Same set of features are extracted for each candidate and her neighbors. $u$ will be mapped to the candidate with most similarity.

We implement and evaluate our proposed attack on two real world datasets, Twitter and Foursquare to study the strengths of anonymization techniques in context of heterogeneous social media data. Our results illustrate that anonymizing even all aspects of data is not sufficient for protecting user privacy due to hidden relations between different aspects of the heterogeneous data. 
\vspace{-15pt}
\section{Future Direction and Advices Sought}
\vspace{-10pt}
Users expect a secure and safe environment when surfing the Web. However, this safety can be jeopardized by threats ranging from advertising trafficking victims~\cite{Alvari2017HT} to having PII exposed to prying eyes. Besides, users' activities on social media which is stored in their browsing history could put their privacy under risk. A recent study has shown the fingerprintability of such data by introducing an attack which maps a given browsing history to a social media profile~\cite{su2017anonymizing}. This identity exposure may result in harms ranging from persecution by governments to targeted frauds~\cite{christin2010dissecting}. Future plans include proposing an effective browsing history anonymizer which benefits from publicly available information of users activities in social media for preserving privacy of users while retaining the utility of web browsing history data. Studying privacy of online users using social media is still at the early stages and consequently there are several challenges ahead. Thus, I seek insights from consortium to overcome the challenges and investigate privacy from either adversarial or defensive perspective.
\vspace{-15pt}
\section{Acknowledgments}
\vspace{-8pt}
This material is an extended abstract of my work published in the 29th ACM conference on Hypertext and Social Media~\cite{htbeigi}. It is supported in part by Army Research Office (ARO) under grant number W911NF-15-1-0328 and Office of Naval Research (ONR) under grant number N00014-17-1-2605.


\vspace{-12pt}
\begin{thebibliography}{}
	\vspace{-8pt}
\scriptsize
\bibitem{htbeigi}
Beigi, G., Shu, K., Zhang, Y. and Liu, H. Securing Social Media User Data - An Adversarial Approach. In the 29th ACM International Conference on Hypertext and Social Media (HT-18).

\bibitem{Beigi2018congruity} Beigi, G., and Liu, H. Similar but Different: Exploiting Users' Congruity for Recommendation Systems. In International Conference on Social Computing, Behavioral-Cultural Modeling, and Prediction. Springer 2018.

\bibitem{narayanan2009anonymizing}
Narayanan, A., and Shmatikov, V. De-anonymizing social networks. In: IEEE Symposium on Security and Privacy 2009.

\bibitem{Alvari2017HT} Alvari, H., Shakarian, P. and Snyder, JEK. Semi-supervised learning for detecting human trafficking. Security Informatics 6, no. 1 (2017): 1.

\bibitem{liu2008towards}
Terzi, E., and Liu, K. Towards identity anonymization on graphs. In: Conference on Management of data 2008.

\bibitem{christin2010dissecting}
Christin, N., Yanagihara, S., and Kamataki, K. Dissecting one click frauds. In: Proceedings of ACM conference on Computer and communications security 2010.

\bibitem{halvarisnam2016} Alvari, H., Hajibagheri, A., Sukthankar, G., and Lakkaraju, K. Identifying community structures in dynamic networks. Social Network Analysis and Mining 6, no. 1 (2016): 77.


\bibitem{TextAnonymization}
	Zhang, J., and Sun, J., and Zhang, R., and Zhang, Y. Privacy-Preserving Social Media Data Outsourcing. In: INFOCOM 2018
\bibitem{su2017anonymizing}
	Su, J., and Shukla, A., and Goel S., and Narayanan, A. De-anonymizing web browsing data with social networks. In: Proceedings of WSDM 2017

\end{thebibliography}
\end{document}